\documentclass[a4paper,11pt]{article}
\usepackage{pos}

\title{High-Energy Neutrino Emission in NGC1068 driven by Turbulent Magnetic Reconnection}
\ShortTitle{Neutrinos in NGC 1068 driven by Turbulent Magnetic Reconnection}

\author*[a]{Luana Passos-Reis}
\author[a]{Elisabete M. de Gouveia Dal Pino}
\author[b]{Juan Carlos Rodríguez-Ramírez}
\author[a]{Giovani H. Vicentin}

\affiliation[a]{Instituto de Astronomia, Geof\'{i}sica e Ci\^{e}ncias Atmosf\'{e}ricas (IAG-USP), Universidade de S\~{a}o Paulo, \\
1226 Rua do Mat\~{a}o, CEP: 05508-090, S\~{a}o Paulo - SP, Brazil.}

\affiliation[b]{Centro Brasileiro de Pesquisas Físicas (CBPF), \\
150 Rua Dr. Xavier Sigaud, CEP: 22290-180, Rio de Janeiro - RJ, Brazil.}

\emailAdd{$^{*}$luana.passos.reis@usp.br}

\abstract{
Astrophysical neutrinos from Active Galactic Nuclei (AGN) offer a unique window into high-energy particle acceleration in obscured environments. The nearby Type II Seyfert galaxy NGC 1068 is a compelling example, exhibiting evidence of a high-energy neutrino excess \cite{IceCube_2022, IceCube_2025} without an associated TeV $\gamma$-ray counterpart. This suggests that hadronic processes may occur within an inner, magnetically dominated region \cite{Murase2022}, where the TeV emission is suppressed by $\gamma \gamma$ absorption and reprocessed via electromagnetic cascades in the dense, obscured environment. Building on our framework, which establishes turbulence-driven magnetic reconnection as the main driver for particle acceleration in this source, we present a refined lepto-hadronic model based on \cite{dalpino_lazarian_2005, kadowaki_etal_15}. In these proceedings, we adopt a conservative inner disk radius compared to our previous results \cite{PassosReis_NGC_ICRC2025, degouveia_PoS_ICRC2025},
moving the acceleration region further from the innermost stable circular orbit.
We estimate the high-energy neutrino emission from hadronic and photo-hadronic processes, constrained by the acceleration timescale for first-order Fermi acceleration within the turbulent current sheet.
The estimated model reproduces the IceCube neutrino flux excess, providing an essential technical complement and validation for our forthcoming comprehensive publication.
}

\FullConference{
High Energy Phenomena in Relativistic Outflows IX (HEPRO-IX)\\
4-8 August 2025\\
Rio de Janeiro, Brazil\\}

\begin{document}
\maketitle




\section{Introduction}

The active galaxy NGC 1068 (Messier 77) has emerged as a multi-messenger source following the detection of a significant high-energy neutrino excess by the IceCube Collaboration \cite{IceCube_2022, IceCube_2025}, 
suggesting the presence of a powerful hadronic component within the source.

However, this detection is notable for the absence of a corresponding TeV $\gamma$-ray counterpart, which is typically produced alongside high-energy neutrinos via hadronic interactions.
This suppression of $\gamma$-rays strongly suggests that the neutrinos originate from a highly obscured environment, such as  the dense accretion disk-corona region \cite{Murase2022}.\footnote{Other regions, such as the torus or the broad-line region, could also be viable production sites.}
In this region, $\gamma$-rays are efficiently absorbed by the ambient photon fields, likely through pair production ($\gamma\gamma$ interactions), due to a high optical depth.

The observed very high-energy (VHE) neutrino excess proposes a fundamental challenge: identifying the physical mechanism(s) capable of efficiently accelerating protons up to the required extreme energies within the inner, magnetized, and dense accretion flow environment.

Recent investigations have explored particle acceleration by plasmoid-driven magnetic reconnection, often requiring additional assumptions about particle pre-acceleration or complex geometries, envolving several combined processes and different acceleration/emission regions for the neutrinos to be produced \cite[e.g.,][and references therein]{Karavola2025,Fiorillo2025}. 
Moreover, plasmoid-mediated reconnection is subject to growth limitations in large-scale MHD flows \cite{Vicentin_etal_2025, Vicentin_etal_2025b} and references therein).

In this work, we investigate an efficient particle acceleration scenario: turbulence-driven magnetic reconnection within the accretion disk-corona. Our model, building upon prior work \cite{dalpino_lazarian_2005, kadowaki_etal_15}, 
assumes that particle acceleration is governed primarily by a first-order Fermi mechanism operating within 
turbulent reconnection layers
(see also \cite{kowal_dgdp_lazarian2012, PassosReis_NGC_ICRC2025, degouveia_PoS_ICRC2025}).
A key advantage of this mechanism is that the acceleration rate is independent of the particle's energy 
(as characteristic of first-order Fermi process at constant reconnection rate \cite{dalpino_lazarian_2005, kowal_dgdp_lazarian2012, kowal_etal_2011,
delvalle2016properties, medina_torrejon_2021,
Medina_Torrejon_2023,
dalpino_tania_2025}). This independence allows for an extremely efficient acceleration that is not limited by the escape time, in contrast to mechanisms where the energy gain grows only linearly with time.

Our present analysis refines previous constraints of \cite{PassosReis_NGC_ICRC2025}, by setting the inner radius further away from the innermost stable circular orbit (ISCO) to circumvent the need for general relativistic corrections\footnote{As we will show, this choice is physically consistent with the assumed magnetic field and density conditions; furthermore, it yields model-dependent parameters that closely align with the source SED.
}, and by treating the reconnection velocity ($v_{\rm rec}$) as a dependent parameter of the model, consistent with earlier formulations \cite{dalpino_lazarian_2005, dalpino_piovezan_2010, kadowaki_etal_15}.

In this paper proceedings, we present preliminary results from this refined model. The full details and main conclusions of this comprehensive analysis will be detailed in a forthcoming work (Passos-Reis et al., in prep.).

\section{Methodology: Core Model and Energetics Analysis}

The framework employs a one-zone model of the NGC 1068 central engine, consisting of a standard geometrically thin, optically thick accretion disk \cite{shakura_sunyaev_73} and an overlying hot, magnetized corona \cite{dalpino_lazarian_2005, dalpino_piovezan_2010, kadowaki_etal_15}. Turbulence within the corona, driven by differential rotation and instabilities such as the magnetorotational instability (MRI) and Parker-Rayleigh-Taylor instability, induces fast magnetic reconnection (\citep{Lazarian_Vishniac_1999, Kadowaki_etal_2018}) within large-scale current sheet, formed where magnetic field lines deposited into  the black hole (BH) horizon \cite{dalpino_lazarian_2005,  MacDonald_etal_1986}
meet those oppositely oriented  arising from the accretion disk.
This process releases  magnetic power, a fraction of which may be channeled into accelerating particles, that will then cascade and generate the observed emission.


The local environmental conditions: specifically the coronal magnetic field ($B_c$), particle number density ($n_c$) and temperature ($T_c$), as well as the disk temperature ($T_d$) at the reconnection region, are consistently derived from a set of coupled equations 
that link the accretion rate ($\dot{m}$), the black hole mass ($m$) and the geometry of the acceleration site (defined by the dimensionless parameters $r_X, l, l_X$). Using the observational constraints for NGC 1068
(see Table \ref{obs_table}), 
these equations provide the physical conditions in the acceleration  region at the corona, 
as derived in \citep{dalpino_lazarian_2005, dalpino_piovezan_2010, kadowaki_etal_15}: 

\begin{align}
\label{eq:full_model_equations}
\left\{
\begin{aligned}
B_{c} &\simeq 9.96 \times 10^{8} r_{X}^{-5/4} \dot{m}^{1/2} m^{-1/2} \text{ G}; \\
\dot{W_{B}} &= 1.66 \times 10^{35} \Gamma^{-1/2} r_{X}^{-5/8} l^{-1/4} l_{X} q^{-2} \dot{m}^{3/4} m; \\
n_{c} &\simeq 8.02 \times 10^{18} \Gamma^{1/2} r_{X}^{-3/8} l^{-3/4} q^{-2} \dot{m}^{1/4} m^{-1} \text{ cm}^{-3}; \\
\Delta R_{X} &\simeq 11.6 \Gamma^{-5/4} r_{X}^{31/16} l^{-5/8} l_{X} q^{-3} \dot{m}^{-5/8} m \text{ cm}; \\
T_{c} &\simeq 2.73 \times 10^{9} \Gamma^{1/4} r_{X}^{-3/16} l^{1/8} q^{-1} \dot{m}^{1/8} \text{ K}; \\
T_{d} &\simeq 3.71 \times 10^{7} \alpha^{-0.25} r_{X}^{-0.37} m^{0.25} \text{ K}.
\end{aligned}
\right.
\end{align}
where $m$ is expressed in units of solar mass ($M_{\odot}$), 
$\dot{m}$, 
in Eddington accretion units ($\dot{M}_{\rm Edd}$), the distance from the  BH to the reconnection site $r_X = R_X / R_{\rm Sch}$, the turbulent reconnection region height $l_X = L_X / R_{\rm Sch}$, and the extension of the corona $l = L / R_{\rm Sch}$, expressed in terms of $R_{\rm Sch}$, the Schwarzschild radius. The factors, $\Gamma = \left(1 + \left( \frac{v_{A0}}{c} \right)^{2} \right)^{-1/2}$ and $q = \left[ 1 - \left(\frac{3 R_{\rm Sch}}{R_{X}} \right)^{1/2} \right]^{1/4}$, account for special-relativistic effects on the Alfvén speed and geometry correction of the inner disk, respectively, where  $v_{A0} = \frac{B}{\sqrt{4 \pi \rho}}$. The model uses the proton rest mass ($m_H$) and a mean molecular weight of $\mu \sim 0.6$, so that $\rho_c = \mu m_{H} n_{c}$.

These equations yield the derived physical parameters presented in Table \ref{table_parameters}.

\begin{table}[htbp]
    \centering
    \caption{Observational Parameters for NGC 1068.}
    \begin{tabular}{ccc}
    \hline
    \hline
    Parameter & Value & Unit \\
    \hline 
        $m$ (Black Hole Mass) & $2 \times 10^7$ & [$M/M_\odot$] \\
        $\dot{m}$ (Accretion Rate) & 0.55 & [$\dot{M}/\dot{M}_{\rm Edd}$] \\
        $d$ (Distance) & $10.1$ & [Mpc] \\
        $z$ (Redshift) & $0.00379$ & \\
    \hline
    \end{tabular}
    \label{obs_table}
\end{table}

\begin{table}[htbp]
    \centering
    \caption{Model-derived parameters, for a fiducial set of free parameters $r_x$, $l_X$, and $l$, adopted for computing cooling rates and the neutrino spectrum of NGC1068.}
    \begin{tabular}{lccc}
    \hline
    \hline
    This Work & Parameter & Value & Unit \\
    \hline
    Coronal Magnetic Flux Tube Height & $l$ & 45.08 & [$L/R_{\rm Sch}$] \\
    Height of Reconnection Region & $l_X$ & 35.06 & [$L_X/R_{\rm Sch}$] \\
    Radial distance from BH & $r_X$ & 6.01 & [$R_X/R_{\rm Sch}$] \\
    \\
    \hline
    \hline
    Model-derived Parameters & & \\
    \hline
    Coronal Magnetic Field & $B_c$ & $1.75 \times 10^{4}$ & [G] \\
    Coronal Particle Density & $n_c$ & $1.57 \times 10^{10}$ & [cm$^{-3}$] \\
    Coronal Temperature & $T_c$ & $3.63 \times 10^{9}$ & [K] \\
    Disk Temperature & $T_d$ & $5.04 \times 10^{5}$ & [K] \\
    Width of Current Sheet & $\Delta R_{X}$ & $1.37 \times 10^{11}$ & [cm] \\
    \quad \quad [in Schwarzschild Radii: $R_{\rm Sch}$] & & $\simeq 0.023$ & [$R_{\rm Sch}$] \\
    Reconnection Power Released & $\dot{W}_{B}$ & $2.05 \times 10^{43}$ & [erg\,s$^{-1}$] \\
    \hline
    \hline
    \end{tabular}
    \label{table_parameters}
\end{table}

The dimensionless free parameters for the present work are constrained to $\mathbf{r_X \simeq 6.0}$ ($R_X/R_{\rm Sch}$), $l \simeq 45$ ($L/R_{\rm Sch}$), and $l_X \simeq 35$ ($L_X/R_{\rm Sch}$),
corresponding to $R_X = 12\, R_g$, $L_X = 70\, R_g$ and $L = 90\, R_g$, where $R_g$ is the gravitational radius. The resulting model-derived parameters, such as the coronal magnetic field $B_c \sim  10^4\, G$ and density $n_c \sim 10^{10}\, \text{cm}^{-3}$, are listed in Table \ref{table_parameters}. We assume a Shakura \& Sunyaev \cite{shakura_sunyaev_73} viscosity parameter $\alpha \simeq 0.1$ for the disk temperature $T_d$.


Our approach, adapted from previous work, allows us to consistently derive the key physical parameters of the corona. It emphasizes first-order Fermi acceleration within the turbulent reconnection sites, as the dominant and highly efficient mechanism. This process leads to energy-independent acceleration 
time, as stressed previously, compared to the energy-dependent drift acceleration time (see \citep{dalpino_tania_2025}, and references therein),
enabling protons to reach the energies required for neutrino production. 

\begin{figure}[h!]
    \centering
    \includegraphics[width=0.9\textwidth]{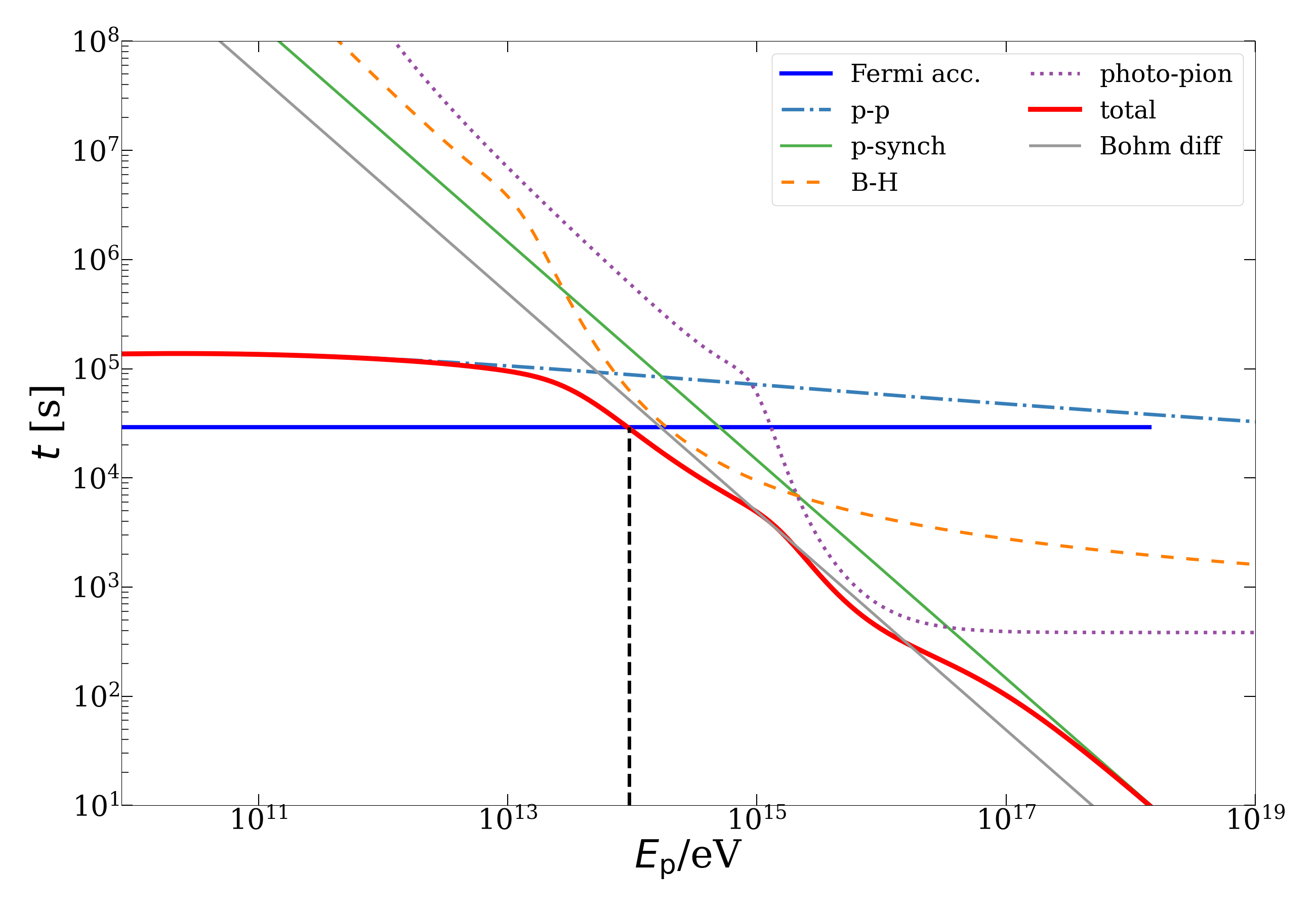}
    \caption{Hadronic Acceleration and Cooling Timescales. This plot illustrates the energetic competition between turbulence-driven reconnection particle acceleration and energy loss mechanisms for protons in the coronal reconnection layer. The first-order Fermi acceleration timescale ($t_{\rm acc}$) is represented by the blue solid horizontal line, confirming its energy-independent nature and high efficiency up to $1.4 \times 10^{18}\, \text{eV}$, where the proton Larmor radius approaches the current sheet width ($\Delta R_X$). Energy loss timescales include synchrotron radiation (green solid line), proton-proton ($t_{\rm pp}$, blue dot-dashed curve), and photo-hadronic interactions ($t_{p\gamma}$, purple dotted curve and Bethe-Heitler, orange dashed curve), which account for both disk blackbody and X-ray photon fields described in details in \cite{PassosReis_NGC_ICRC2025}. The maximum proton energy achievable ($E_{\max}$) is determined by the intersection point where $t_{\rm acc}$ equals the sum of the total losses in our system $\sim 10^{14}\, \text{eV}$.}
    \label{fig:cool_HAD}
\end{figure}

Specifically, the acceleration time due to first-order Fermi within a turbulent current sheet is given by (\citep{Xu_Lazarian_2022, dalpino_tania_2025}; see also \cite{degouveia_PoS_ICRC2025}):

\begin{equation}
    t_{\rm acc} \simeq \frac{4 \Delta R_X}{c d},
\end{equation}

where $d \approx \frac{2\beta_{\rm in}(3\beta_{\rm in}^2+3\beta_{\rm in}+1)}{3(\beta_{\rm in}+0.5)(1-\beta_{\rm in}^2)}$,
with a reconnection velocity 
$\beta_{\rm in}=v_{\rm rec}/c \sim 0.001$, 
given our model constraints and $v_{\rm rec} = \Delta R_{X}\ v_{A} / L_{X}$ \footnote{We note that, although this reconnection velocity is lower than the values typically reported in turbulence-driven reconnection simulations \cite[e.g.][]{kowal_dgdp_lazarian2012, delvalle2016properties, kadowaki_etal_2021}, it is consistent with the reconnection-layer geometrical parameters inferred for the inner disk/coronal region of NGC~1068, and large enough   to reproduce the observed SED.}.
The interplay between the acceleration and cooling mechanisms, shown in Fig.~\ref{fig:cool_HAD}, defines the maximum achievable proton energy as the intersection of the acceleration and total energy loss curves.




The magnetic recconection power, $\dot{W_{B}}$, which is released by turbulence-driven fast reconnection in the surrounds of the BH is essential for characterizing the efficiency of particle acceleration. 
A fraction of the magnetic power released ($\sim 80\%$) is assumed to be injected into the accelerated protons with a power-law index of $1.6$,
modeled as a function of the particle number density and the interaction timescale.\footnote{We note that laboratory experiments and observations of solar flares suggest that roughly 50-60\% of the magnetic reconnection power is channeled into particle acceleration \citep{yamada_etal_2014}. Our adoption of 80\% is therefore somewhat higher. Nevertheless, in the paper in preparation we explore a broader parameter space in which this assumption is relaxed and smaller fractions are also considered. Regarding the power-law index (p=1.6), this value is consistent with results from numerical studies of reconnection-driven acceleration, including both PIC simulations and MHD simulations with test particles (e.g., \cite{guo_2014, sironi_spitkovsky_2014, kowal_dgdp_lazarian2012, medina_torrejon_2021, Medina_Torrejon_2023, dalpino_tania_2025} and references therein).
}
These protons will undergo subsequent hadronic interactions, primarily proton-proton (p-p), which produce high-energy neutrinos and gamma-rays through the decay of pions ($\pi^{0}$ and $\pi^{\pm }$).

For the photo-hadronic processes, the ambient photon field is modeled using two components as in \cite{PassosReis_NGC_ICRC2025}: a disk blackbody radiation (serving as an Optical-UV proxy) and an X-ray component constrained by observations \cite{Bauer2015, Marinucci2016}. The high optical depth created by these two radiation fields ensures the efficient absorption of $\gamma$-rays via pair production ($\gamma\gamma$ annihilation), 
providing a self-consistent explanation for the VHE neutrino excess without a corresponding electromagnetic $\gamma$-ray counterpart, as shown in \cite{PassosReis_NGC_ICRC2025}. The corresponding $\tau_{\gamma \gamma}$ diagram is presented in the aforementioned paper in preparation.


\section{Results \& Discussion: High-Energy Neutrino Spectrum}

\begin{figure}[h!]
    \centering
    \includegraphics[width=0.8\textwidth]{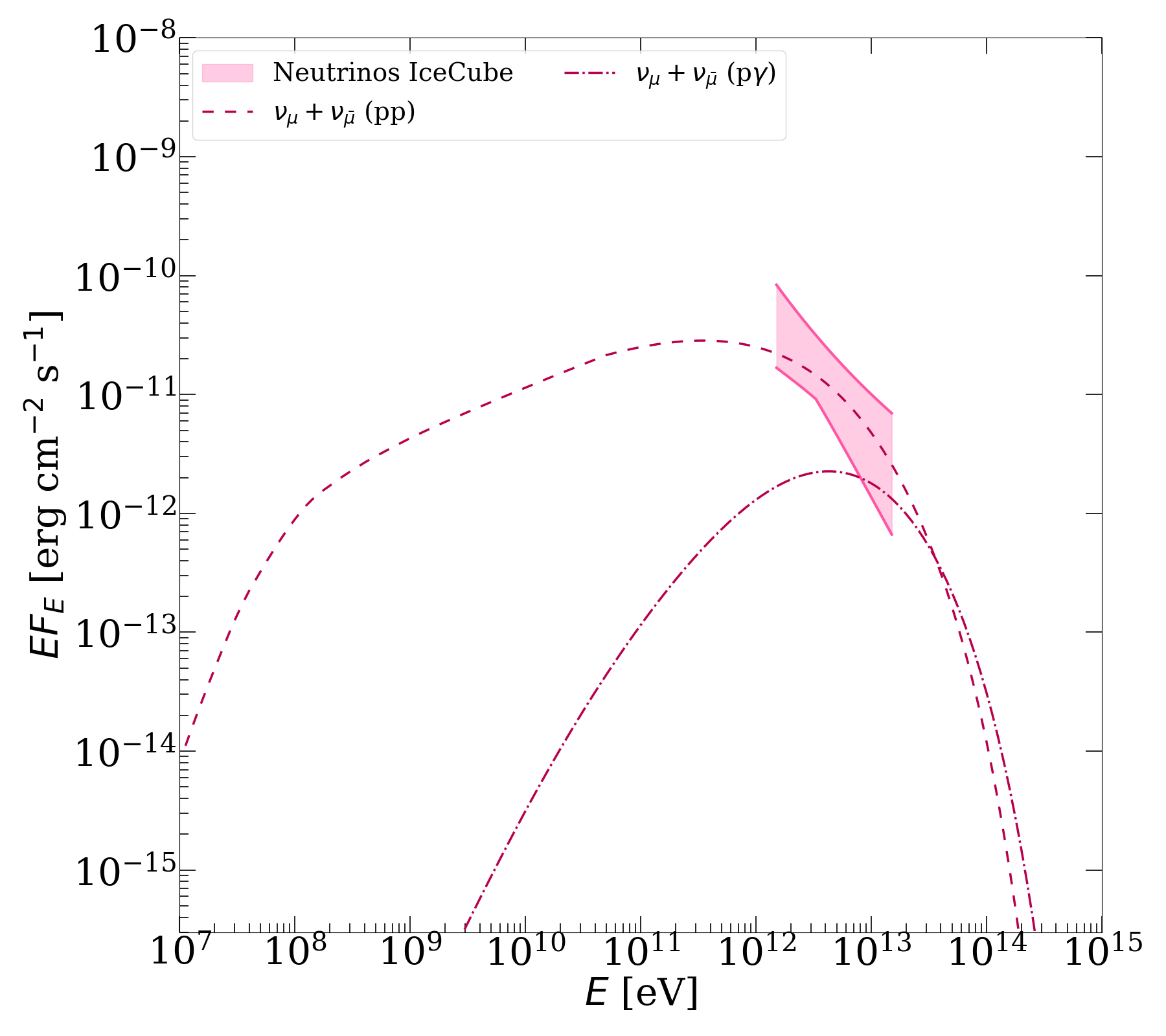}
    \caption{High-energy neutrino emission modeled by the cascading of hadrons. The acceleration of protons up to the energies seen in Fig.~\ref{fig:cool_HAD} allows protons to reach the energies required for subsequent hadronic and photo-hadronic interactions (cascading). These interactions produce the observed high-energy neutrinos modeled in this plot, constrained by $R_X \simeq 6\, R_{\rm Sch}$. The model successfully reproduces the observed IceCube neutrino flux 95\% confidence region \cite{IceCube_2022} (shaded region in magenta), confirming the efficient hadronic acceleration by turbulent magnetic reconnection.
    }
    \label{fig:neutrino_bumps}
\end{figure}

The refined lepto-hadronic model, constrained by a conservative inner disk radius of $R_X \simeq 6\, R_{\rm Sch}$, successfully reproduces the IceCube neutrino flux 95\% confidence region (magenta shaded region in Fig.~\ref{fig:neutrino_bumps}). We demonstrate the dominance of first-order Fermi acceleration within the large scale turbulent current sheet. 
This process allows protons to reach a maximum energy of $E_{\max} \sim 10^{14}\, \text{eV}$, which is compatible with IceCube neutrino production, as we can see from  Figure \ref{fig:neutrino_bumps} (see also \cite{PassosReis_NGC_ICRC2025}),  
and sufficient to drive the subsequent high-energy emission.

Our work validates the model of turbulence-driven magnetic reconnection as a strong candidate mechanism for high-energy particle acceleration in NGC 1068, 
providing a physical explanation for the observed IceCube neutrino excess.

The modeled high-energy neutrino curves in Figure \ref{fig:neutrino_bumps} are the result of hadronic and photo-hadronic interactions driven by the accelerated protons, that will subsequently cascade generating the high-energy emission. Specifically, they originate from proton-proton (p-p) interactions and  photo-hadronic interactions ($p\gamma$ interactions), primarily photo-pion cascading, which involve the protons interacting with the disk blackbody radiation (Optical-UV) and X-ray photon fields. The particle interactions were approximated through timescale estimates, following the first-order Fermi acceleration and cooling mechanisms described in detail in the recent work \cite{dalpino2024}.

As shown in \cite{PassosReis_NGC_ICRC2025}, the resulting $\gamma$-rays, produced alongside neutrinos via neutral pion decay ($\pi^0 \rightarrow \gamma\gamma$), are efficiently absorbed 
within the dense accretion disk-corona environment through pair production ($\gamma\gamma$ annihilation). This mechanism creates a high optical depth, which provides a self-consistent explanation for the absence of a corresponding TeV $\gamma$-ray counterpart.

This proceedings paper is a critical step toward our forthcoming comprehensive publication (Passos Reis et al., in prep.), which will present a full treatment of the lepto-hadronic cascading and the resulting multi-messenger spectrum, while exploring a broader range of values for the model’s free parameters.

This work has received partial support from FAPESP (grants 2021/02120-0 \& 2024/05459-6).



\begin{thebibliography}{99}

\bibitem{IceCube_2025}
Abbasi, R. et al. (2025). IceCube Collaboration.
\textit{arXiv e-prints}, \textbf{2510.13403}. 
\href{https://arxiv.org/abs/2510.13403}{arXiv:2510.13403}


\bibitem{IceCube_2022}
Abbasi, R. et al. (2022). IceCube Collaboration.
\textit{Science}, \textbf{378(6619)}, 538--543.
\href{https://doi.org/10.1126/science.abg3395}{doi:10.1126/science.abg3395}


\bibitem{Murase2022}
Murase, K. (2022).
\textit{ApJL}, \textbf{941}, L17. \href{https://doi.org/10.3847/2041-8213/aca53c}{doi:10.3847/2041-8213/aca53c}


\bibitem{dalpino_lazarian_2005}
de Gouveia Dal Pino, E. M. \& Lazarian, A. (2005).
\textit{A\&A}, \textbf{441(3)}, 845–853. \href{https://doi.org/10.1051/0004-6361:20042590}{doi:10.1051/0004-6361:20042590}


\bibitem{kadowaki_etal_15}
Kadowaki, L.~H.~S., de Gouveia Dal Pino, E.~M., \& Singh, C.~B. (2015).
\textit{ApJ}, \textbf{802}, 113.  
\href{https://doi.org/10.1088/0004-637X/802/2/113}{doi:10.1088/0004-637X/802/2/113}



\bibitem{PassosReis_NGC_ICRC2025}
Passos Reis, L., de Gouveia Dal Pino, E. M., Rodríguez Ramírez, J. C. \& Vicentin, G. H. (2025).
\textit{PoS(ICRC2025)}, \textbf{501}, 1143. \href{https://doi.org/10.22323/1.501.1143}{doi:10.22323/1.501.1143}


\bibitem{degouveia_PoS_ICRC2025}
de Gouveia Dal Pino, E., Kadowaki, L. H., Medina-Torrejon, T. E. et al. (2025).
\textit{PoS(ICRC2025)}, \textbf{501}, 032. \href{https://doi.org/10.22323/1.501.0032}{doi:10.22323/1.501.0032}


\bibitem{Fiorillo2025}
Fiorillo, D. F. G., Petropoulou, M., Comisso, L., Peretti, E. \& Sironi, L. (2024).
\textit{ApJL}, \textbf{961}, L14. \href{https://doi.org/10.3847/2041-8213/ad192b}{doi:10.3847/2041-8213/ad192b}


\bibitem{Karavola2025}
Karavola, D., Petropoulou, M., Fiorillo, D. F. G., Comisso, L. \& Sironi, L. (2025).
\textit{JCAP}, \textbf{2025(04)}, 075. \href{https://doi.org/10.1088/1475-7516/2025/04/075}{doi:10.1088/1475-7516/2025/04/075}


\bibitem{Vicentin_etal_2025}
Vicentin, G. H., Kowal, G., de Gouveia Dal Pino, E. M., \& Lazarian, A. (2025).
\textit{ApJ}, \textbf{987}, 213. \href{https://doi.org/10.3847/1538-4357/addc62}{doi:10.3847/1538-4357/addc62}


\bibitem{Vicentin_etal_2025b}
Vicentin, G. H., Kowal, G., de Gouveia Dal Pino, E. M. \& Lazarian, A. (2025).
\textit{arXiv e-prints}, \textbf{2510.01060}. \href{https://arxiv.org/abs/2510.01060}{arXiv:2510.01060}


\bibitem{kowal_dgdp_lazarian2012}
Kowal, G., de Gouveia Dal Pino, E. M. \& Lazarian, A. (2012).
\textit{Phys. Rev. Lett.}, \textbf{108}, 241102. \href{https://doi.org/10.1103/PhysRevLett.108.241102}{doi:10.1103/PhysRevLett.108.241102}


\bibitem{kowal_etal_2011}
Kowal, G., de Gouveia Dal Pino, E. M. \& Lazarian, A. (2011). \textit{ApJ}, \textbf{735}, 102. \href{https://doi.org/10.1088/0004-637X/735/2/102}{doi:10.1088/0004-637X/735/2/102}


\bibitem{Medina_Torrejon_2023}
Medina-Torrejón, T. E., de Gouveia Dal Pino, E. M. \& Kowal, G. (2023). \textit{ApJ}, \textbf{952}, 168. \href{https://doi.org/10.3847/1538-4357/acd699}{doi:10.3847/1538-4357/acd699}


\bibitem{delvalle2016properties}
del Valle, M. V., de Gouveia Dal Pino, E. M. \& Kowal, G. (2016).
\textit{MNRAS}, \textbf{463(4)}, 4331--4343. \href{https://doi.org/10.1093/mnras/stw2276}{doi:10.1093/mnras/stw2276}

\bibitem{medina_torrejon_2021}
Medina-Torrejón, T. E., de Gouveia Dal Pino, E. M., Kadowaki et al. (2021).
\textit{ApJ}, \textbf{908}, 193. \href{https://doi.org/10.3847/1538-4357/abd6c2}{doi:10.3847/1538-4357/abd6c2}

\bibitem{dalpino_tania_2025}
de Gouveia Dal Pino, E. M. \& Medina-Torrejon, T. E. (2025).
\textit{arXiv e-prints}, \textbf{2410.13071}. \href{https://arxiv.org/abs/2410.13071}{arXiv:2410.13071}


\bibitem{dalpino_piovezan_2010}
de Gouveia Dal Pino, E. M., Piovezan, P. P. \& Kadowaki, L. H. S. (2010).
\textit{A\&A}, \textbf{518}, A5. \href{https://doi.org/10.1051/0004-6361/200913462}{doi:10.1051/0004-6361/200913462}


\bibitem{shakura_sunyaev_73}
Shakura, N.~I., \& Sunyaev, R.~A. (1973).  
\textit{A\&A}, \textbf{24}, 337.
\href{https://adsabs.harvard.edu/full/1973A%26A....24..337S}{ADS:1973A\&A....24..337S}

\bibitem{Lazarian_Vishniac_1999}
Lazarian, A. \& Vishniac, E. T. (1999).
\textit{ApJ}, \textbf{517}, 700. \href{https://doi.org/10.1086/307233}{doi:10.1086/307233}


\bibitem{Kadowaki_etal_2018}
Kadowaki, L. H. S., de Gouveia Dal Pino, E. M., \& Stone, J. M. (2018).
\textit{ApJ}, \textbf{864}, 52. \href{https://doi.org/10.3847/1538-4357/aad4ff}{doi:10.3847/1538-4357/aad4ff}




\bibitem{MacDonald_etal_1986}
MacDonald, D. A., Thorne, K. S., Price, R. H. \& Zhang, X.-H. (1986).
In: Thorne, K. S., Price, R. H. \& MacDonald, D. A. (1986), \textit{Black Holes: The Membrane Paradigm (A87-28141 11-90)}, \textit{Yale University Press}, pp. 121--145. \href{https://ui.adsabs.harvard.edu/abs/1986bhmp.book..121M}{ADS:1986bhmp.book..121M}

\bibitem{Xu_Lazarian_2022}
Xu, S. \& Lazarian, A. (2023).
\textit{ApJ}, \textbf{942}, 21. \href{https://doi.org/10.3847/1538-4357/aca32c}{doi:10.3847/1538-4357/aca32c}



\bibitem{kadowaki_etal_2021}
Kadowaki, L. H. S., de Gouveia Dal Pino, E. M., Medina-Torrejón, T. E., Mizuno, Y. \& Kushwaha, P. (2021).
\textit{ApJ}, \textbf{912}, 109. \href{https://doi.org/10.3847/1538-4357/abee7a}{doi:10.3847/1538-4357/abee7a}



\bibitem{Marinucci2016}
Marinucci, A., Bianchi, S., Matt, G., et al. (2016).
\textit{MNRASL}, \textbf{456(1)}, L94--L98. \href{https://doi.org/10.1093/mnrasl/slv178}{doi:10.1093/mnrasl/slv178}


\bibitem{Bauer2015}
Bauer, F.~E. et al. (2015).
\textit{ApJ}, \textbf{812}, 116. \href{https://doi.org/10.1088/0004-637X/812/2/116}{doi:10.1088/0004-637X/812/2/116}


\bibitem{yamada_etal_2014}
Yamada, M., Yoo, J., Jara-Almonte, J. et al. (2014).
\textit{Nat. Commun.}, \textbf{5}, 4774. \href{https://doi.org/10.1038/ncomms5774}{doi:10.1038/ncomms5774}


\bibitem{sironi_spitkovsky_2014}
Sironi, L., \& Spitkovsky, A. (2014). \textit{ApJL}, \textbf{783}, L21. \href{https://doi.org/10.1088/2041-8205/783/1/L21}{doi:10.1088/2041-8205/783/1/L21}


\bibitem{guo_2014}
Guo, F., Li, H., Daughton, W.,  Liu, Y.-H. (2014). \textit{Phys. Rev. Lett.}, \textbf{113}, 155005. \href{https://doi.org/10.1103/PhysRevLett.113.155005}{doi:10.1103/PhysRevLett.113.155005}



\bibitem{dalpino2024}
de Gouveia Dal Pino, E.~M., Rodríguez-Ramírez, J.~C., \& del Valle, M.~V. (2025).  
\textit{MNRAS}, \textbf{537(4)}, 3895--3907. \href{https://doi.org/10.1093/mnras/staf251}{doi:10.1093/mnras/staf251}







\end{thebibliography}
\end{document}